\documentclass[conference]{IEEEtran}

 \IEEEoverridecommandlockouts

\usepackage{bm}
\usepackage{algorithmic,algorithm}
\usepackage{cite}
\usepackage{amsmath,amssymb,amsfonts}
\usepackage{algorithmic}
\usepackage{graphicx}
\usepackage{textcomp}
\usepackage{xcolor}
\def\BibTeX{{\rm B\kern-.05em{\sc i\kern-.025em b}\kern-.08em
        T\kern-.1667em\lower.7ex\hbox{E}\kern-.125emX}}

\begin{document}

\title{\huge Age-of-Information-based Scheduling in Multiuser Uplinks with Stochastic Arrivals: A POMDP Approach
\thanks{The work of H. Chen is supported by the CUHK direct grant under the project code 4055126.}
    \thanks{The first two authors contributed equally to this paper.  Any technical problems of this paper should go to H. Chen. }
}
\author{\IEEEauthorblockN{Aoyu Gong\textsuperscript{$*$}, Tong Zhang\textsuperscript{$\dagger$}, He Chen\textsuperscript{$\dagger$}, and Yijin Zhang\textsuperscript{$*$}
    }%

    \IEEEauthorblockA{\textsuperscript{$*$}School of Electronic and Optical Engineering, Nanjing University of Science and Technology, Nanjing 210094, China}
\IEEEauthorblockA{\textsuperscript{$\dagger$}Department of Information Engineering, The Chinese University of Hong Kong, Hong Kong SAR, China}
     Email:
     gongaoyu@gmail.com,
     bennyzhangtong@yahoo.com,
     he.chen@ie.cuhk.edu.hk,
     yijin.zhang@gmail.com
}

\maketitle

\begin{abstract}
 In this paper, we consider a multiuser uplink status update system, where a monitor aims to timely collect randomly generated status updates from multiple end nodes through a shared wireless channel. We adopt the recently proposed metric, termed age of information (AoI), to quantify the information timeliness and freshness. Due to the random generation of the status updates at the end node side, the monitor only grasps a partial knowledge of the status update arrivals. Under such a practical scenario, we aim to address a fundamental multiuser scheduling problem: how to schedule the end nodes to minimize the network-wide AoI? To solve this problem, we formulate it as a partially observable Markov decision process (POMDP), and develop a dynamic programming (DP) algorithm to obtain the optimal scheduling policy. By noting that the optimal policy is computationally prohibitive, we further design a low-complexity myopic policy that only minimizes the one-step expected reward. Simulation results show that the performance of the myopic policy can approach that of the optimal policy, and is better than that of the baseline policy.
\end{abstract}

\section{Introduction}
The information freshness has become an increasingly important performance metric in this era of the Internet of Things (IoT). Various IoT services, such as remote monitoring and control, require the underlying information to be delivered as timely as possible \cite{000,001}. To quantify the information timeliness and freshness, the age of information (AoI) metric, defined as the time elapsed since the generation time of the latest received status update at the monitor, has been investigated in \cite{-1,1,2,3,4}. Early work (e.g., \cite{1,2,3,4,gu2019timely,gu2019Minimizing,wang2019minimizing2}) on the AoI focused on single-user systems, while recent work (e.g., \cite{5,wang2020minimizing,8,9,11,12,13}) shifted to multi-user systems, such as broadcast systems and multiuser uplink systems, where the AoI not only depends on the single-user behaviors but also depends on the interactions among different end nodes.

In broadcast systems, scheduling problems of minimizing the network-wide AoI were studied in \cite{5,wang2020minimizing,8,9}. The authors in \cite{5} considered the ``generate-at-will" model for the status update, where the status update arrivals could be generated by end nodes once they were scheduled to transmit. Three low-complexity scheduling policies were developed and analyzed in \cite{5}, including a randomized policy, a max-weight policy and a Whittle'¡¯s index policy. The authors in \cite{wang2020minimizing} extended the work in \cite{5} by studied the nonorthogonal multiple access. Considering event-triggered measurements where the status update arrivals are stochastic, the authors in \cite{8} derived a universal lower bound of scheduling policies. In \cite{9}, both ``generate-at-will" and stochastic arrival models with no buffer at end nodes were investigated, and an Whittle's index policy was proposed to achieve the performance close to a structural Markov decision process (MDP) algorithm.

In multiuser uplink systems, scheduling problems of minimizing the network-wide AoI is more challenging than that in broadcast systems, especially when the status update arrivals are stochastic. This is mainly because the monitor may not know whether new status updates arrive at end nodes. Most existing work assumed end nodes used extra feedback overhead to report their status update arrivals so that the monitor had a complete knowledge of their status update arrivals \cite{11, 12}. Such feedback leads to considerable overhead and thus makes the corresponding scheduling policies hard to implement in practice.

To combat this weakness, we consider a multiuser uplink system with stochastic status update arrivals. We assume that there is no extra feedback overhead for end nodes to report their status update arrivals. Thus, the monitor can obtain the status update arrival knowledge of an end node only when it is scheduled to transmit and its transmission is successful. Such a practical assumption leads to a partial knowledge of status update arrivals at the monitor. In this context, we aim to minimize the expected weighted-sum AoI (EWSAoI) of all end nodes by designing multiuser scheduling policies. Note that the consideration of a partial knowledge of status update arrivals renders difficulties in solving the scheduling policies in the considered system.

The main contributions of this paper are summarized as follows. We formulate the considered scheduling problem as a partially observable Markov decision process (POMDP), of which the belief state characterizes the fully observable AoI and the partially observable status update arrivals of end nodes at the monitor. Built on this POMDP, we develop a dynamic programming (DP) algorithm to solve the optimal policy. To reduce the computational complexity, we further propose a low-complexity myopic policy that only minimizes the one-step expected reward. Simulation results show that the performance of the myopic policy is very close to that of the optimal policy solved by the DP algorithm. Both of them are superior to the baseline policy utilizing no knowledge of status update arrivals. To the best of our knowledge, this is the first work that designs an information-freshness-oriented multiuser scheduling policy under partial system information.

\section{System Model and Problem Formulation}

In this section, we first describe the system model, and then formulate the network-wide AoI minimization problem.

\subsection{System Model}

As shown in Fig. 1, we consider a multiuser uplink system where $K$ end nodes report their freshest (i.e. most recently arriving) status updates to a common monitor via a shared channel. The $K$ end nodes are identified by the index set $\bm{{\cal{K}}} \triangleq \{1,2,\ldots,K\}$. The time axis is divided into time slots of equal duration. We let $T$ denote the time-horizon of the discrete-time system considered. In each time slot $t \in \{ 1,2,\ldots,T \}$, a new status update arrives at end node $i \in \bm{{\cal{K}}}$ according to a Bernoulli arrival process with mean $\lambda_i \in (0,1]$. The arrival process is independent and identically distributed over time, and independent across end nodes. At the beginning of each time slot, the monitor will schedule an end node to transmit its freshest status update. The transmissions of end nodes to the monitor are error-prone. Specifically, the transmission of end node $i$ has a successful probability $p_i$, and an error probability $(1 - p_i)$.

\begin{figure}[!ht]
    \centering
    \includegraphics[width=2.8in]{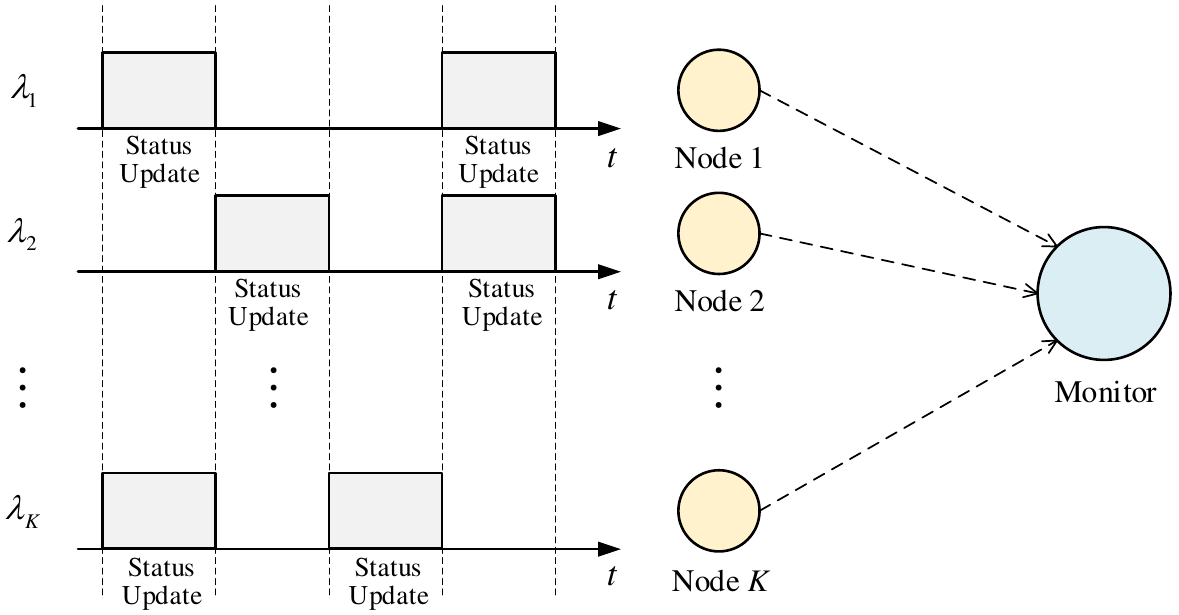}
    \caption{The multiuser uplink system with stochastic arrivals of status updates.}
\end{figure}

\subsection{Local Age}

Each end node is assumed to store at most one status update in the buffer. When a new status update arrives at an end node, the end node drops the status updates already in its buffer if its buffer is not empty. This assumption ensures one status update stored in the buffer of an end node is freshest. The local age of end node $i$, denoted by $z_i^t$, measures the freshness of the status update at the end node. The evolution of $z_i^t$ can be expressed as follows:
\begin{equation}
z_i^{t+1} =
\begin{cases}
z_i^t + 1, &\text{No arrival in time slot $t$,} \\
1, &\text{Status update arrival in time slot $t$.}
\end{cases} \label{ST}
\end{equation}
As shown in \eqref{ST}, the local age of end node $i$ will increase by $1$ if there is no status update arrival and be reset to 1 otherwise.
The local age of different end nodes evolves independently according to their Bernoulli arrival processes. Without loss of generality, we assume that there is a status update arrival at the beginning of the first time slot for each end node.

In each time slot, reporting the local age of all end nodes to the monitor causes a large amount of extra overhead.
For practical implementations, we enforce that the local age of an end node can be observed by the monitor only when the end node is scheduled and successfully transmits its freshest status update. This is because the status update received by the monitor contains the time-stamp of itself.

\subsection{AoI Minimization with Partial Knowledge of Arrivals}

In this paper, we adopt the AoI metric to quantify the information freshness.
The AoI of end node $i$ at the monitor, denoted by $h_i^t$, will be set to the local age of end node $i$, if the end node is scheduled and its transmission is successful. Otherwise, the AoI of the end node will increase by 1. The AoI evolution can be expressed as follows:
\begin{equation}
h_i^{t+1} = \begin{cases}
z_i^{t} + 1, & \text{Scheduled and received,} \\
h_i^{t} + 1, & \text{Otherwise.}
\end{cases}
\end{equation}
Note that, since the local age of an end node increases when there is no new status update arrival, the monitor schedules the end node with no status update stored in its buffer will not reduce its AoI at the monitor.

In this context, the monitor is only aware of the local age of an end node that is scheduled and transmits successfully. This leads to partial observation of the system information at the monitor. With such partial knowledge, we aim to find a scheduling policy $\bm{\pi}$ minimizing the EWSAoI, which can be formulated as the following optimization problem:
\begin{equation} \label{EWSAoI}
(\text{P1}):\quad \min_{\bm{\pi}} \,\, \frac{1}{TK} \mathbb{E}\left[\left. {\sum_{t=1}^T \sum_{i=1}^K \omega_i h_i^{t}}\right|\bm{\pi} \right],
\end{equation}
where $\omega_i \in (0,{\cal{1}})$ is the importance weight of end node $i$. The expectation is taken over all system dynamics.

\section{POMDP Formulation}

In this section, to solve the problem (P1), we reformulate it as a POMDP, and use the average reward of the POMDP to evaluate the EWSAoI.
\begin{enumerate}
    \item \underline{\emph{States:}}
    We denote the state of end node $i$ in time slot $t$ by $\textbf{s}_i^t \triangleq[h_i^t,z_i^t]$, where $h_i^t \in  \bm{{\cal{T}}} \triangleq \{1,2,3,\ldots\}$ is its instantaneous AoI at the monitor and $z_i^t \in \bm{{\cal{T}}}$ is its local age.
    Then, we denote the state of the POMDP in time slot $t$ by $\textbf{s}^t \triangleq[\textbf{h}^t,\textbf{z}^t]$, where $\textbf{h}^t\triangleq [h_1^t,h_2^t,\ldots,h_K^t] \in \bm{{\cal{H}}} \triangleq \bm{{\cal{T}}}^K$ represents the AoI of all end nodes, and $\textbf{z}^t \triangleq [z_1^t,z_2^t,\ldots,z_K^t] \in \bm{{\cal{Z}}} \triangleq \bm{{\cal{T}}}^K$ represents the local age of all end nodes.
    Denote by $\bm{\mathcal{S}}$ the space of all possible states.

    \item \underline{\emph{Actions:}}
    We denote the action of the POMDP in time slot $t$ by $\textbf{a}^t \triangleq [a_1^t,a_2^t,\ldots,a_K^t]$, where $a_i^t \in {\cal{A}} \triangleq \{0,1\}$  indicates whether end node $i$ is scheduled to transmit or not in time slot $t$. If end node $i$ is scheduled, then $a_i^t = 1$; otherwise, $a_i^t = 0$. In the single-antenna system considered, the monitor can only schedule at most one end node in each time slot.
    Thus, we have $\sum_{i=1}^K a_i^t \leq 1$.
    Denote by $\bm{\mathcal{A}}$  the space of all possible actions.

    \item \underline{\emph{Observations:}} The observations of the POMDP at the monitor consists of the fully observed AoI and partially observed local age of all end nodes.
    Specifically, if end node $i$ is scheduled and its transmission is successful, its local age can be accurately observed by the monitor. Otherwise, there is no observation of its local age. We denote the observation of the POMDP in time slot $t$ by $\textbf{o}^t \triangleq [\textbf{o}_1^1,\textbf{o}_2^t,\ldots,\textbf{o}_K^t]$, where $\textbf{o}_i^t \triangleq [h_i^t,\hat{z}_i^t]$ is the observation of end node $i$, including its fully observed AoI $h_i^t$ and partially observed local age $\hat{z}_i^t \in \{\bm{{\cal{T}}},X\}$.
    Note that $X$ means no observation of the local age of an end node, caused by its unsuccessful transmission or not being scheduled.
    Denote by $\bm{\mathcal{O}}$  the space of all possible observations.

    \item \underline{\emph{Belief States:}}
    We denote the belief state of the POMDP in time slot $t$ by $\textbf{I}^t \triangleq [\textbf{h}^t, \textbf{b}^t]$, where $\textbf{b}^t$ is a probability distribution over $\bm{\mathcal{Z}}$.
    Let $b^t(\textbf{z}^t)$ denote the probability assigned to state $\textbf{z}^t$ by distribution $\textbf{b}^t$, which satisfies $b^t(\textbf{z}^t) \in [0,1]$ for all $\textbf{z}^t \in \bm{\mathcal{Z}}$, and $\sum_{\textbf{z}^t \in \bm{\mathcal{Z}}} b^t(\textbf{z}^t)=1$.
    It is worth mentioning that although in general the belief state of a POMDP is a probability distribution over $\bm{\mathcal{S}}$, in our problem $\textbf{h}^t$ is fully observable, i.e., its belief state update is always deterministic given $\textbf{h}^{t-1}$, $\textbf{a}^{t-1}$ and $\textbf{o}^{t-1}$.
    Denote by $\bm{\mathcal{I}}$ the space of all possible belief states.

    \item  \underline{\emph{State Transition Function and Observation Function:}} Because the belief state update of $\textbf{h}^t$ is deterministic, we only need to define the state transition function and observation function of $\textbf{z}^t$. The state transition function is denoted by $\text{Pr}(\textbf{z}^{t+1}|\textbf{z}^{t})=\prod_{i=1}^K\text{Pr}(z_i^{t+1}|z_i^{t})$, giving the conditional probability of reaching state $\textbf{z}^{t+1}$ given state $\textbf{z}^t$. For end node $i \in \bm{{\cal{K}}}$, we have
    \begin{equation}
    \text{Pr}(z_i^{t+1}|z_i^{t}) =
    \begin{cases}
    \lambda_i, & \text{if}\,\,z_i^{t+1} = 1, \\
    1-\lambda_i, & \text{if}\,\,z_i^{t+1} = z_i^{t}+1, \\
    0, & \text{otherwise}.
    \end{cases}
    \end{equation}
    The observation function is denoted by $\text{Pr}(\textbf{o}^{t}|\textbf{z}^{t},\textbf{a}^{t})=\prod_{i=1}^K \text{Pr}(\textbf{o}_i^{t}|z_i^{t},a_i^{t})$, giving the conditional probability of making observation $\textbf{o}^{t}$ given state $\textbf{z}^{t}$ and action $\textbf{a}^t$.
    If end node $i$ is scheduled, then
    \begin{equation}
    \text{Pr}(\textbf{o}_i^{t}|z_i^{t},1) =
    \begin{cases}
    p_i, & \text{if}\,\,\hat{z}_i^{t} = z_i^{t}, \\
    1-p_i, & \text{if}\,\,\hat{z}_i^{t} = X, \\
    0, & \text{otherwise}.
    \end{cases}
    \end{equation}
    If end node $i$ is not scheduled, then $\text{Pr}(\textbf{o}_i^{t}|z_i^{t},0) = 1$ if $\hat{z}_i^{t} = X$, and $\text{Pr}(\textbf{o}_i^{t}|z_i^{t},0) = 0$ otherwise.

    \item \underline{\emph{Belief State Update:}} In our POMDP, the monitor keeps belief states rather than knowing actual states.
    In time slot $t$, belief state $\textbf{I}^t$ is a sufficient statistic for a given history $\{\textbf{I}^1,\textbf{a}^1,\textbf{o}^1,\textbf{a}^2,\textbf{o}^2,\ldots,\textbf{a}^{t-1},\textbf{o}^{t-1}\}$, consisting of two parts: $\textbf{h}^t$ and $\textbf{b}^t$.
    When given $\textbf{I}^t$, $\textbf{a}^t$ and $\textbf{o}^t$, for $\forall i \in \bm{{\cal{K}}}$, $h_i^{t+1}$ can be updated as follows:
    \begin{equation} \label{eq:h-update}
    h_i^{t+1} =
    \begin{cases}
    \hat{z}_i^t+1, & \text{if}\,\,\hat{z}_i^{t} \neq X,  \\
    h_i^t+1, & \text{if}\,\,\hat{z}_i^{t} = X. \\ 
    \end{cases}
    \end{equation}
    As shown in  \eqref{eq:h-update}, the update of $\textbf{h}^{t}$ is always deterministic.
    When given the same condition, for $\forall \textbf{z}^{t+1} \in \bm{{\cal{Z}}}$, $b^{t+1}(\textbf{z}^{t+1})$ can be updated via the Bayes' theorem:
    \begin{equation} \label{eq:b-update}
    b^{t+1}(\textbf{z}^{t+1}) = \eta \sum_{\textbf{z}^{t} \in \bm{\mathcal{Z}}} \text{Pr} (\textbf{o}^t|\textbf{z}^t,\textbf{a}^t) \text{Pr} (\textbf{z}^{t+1}|\textbf{z}^t) b^{t}(\textbf{z}^{t}),
    \end{equation}
    where
    \[
    \eta = 1/\sum_{\textbf{z}^{t+1} \in \bm{\mathcal{Z}}} \sum_{\textbf{z}^{t} \in \bm{\mathcal{Z}}} \{\text{Pr}(\textbf{o}^t|\textbf{z}^t,\textbf{a}^t)\text{Pr} (\textbf{z}^{t+1}|\textbf{z}^t) b^{t}(\textbf{z}^{t})\}
    \]
    is a normalizing factor.
    We denote the update of $\textbf{h}^{t}$ in \eqref{eq:h-update} and the update of $\textbf{b}^t$ in \eqref{eq:b-update} by the update function $\textbf{I}^{t+1} = f(\textbf{I}^t,\textbf{a}^t,\textbf{o}^t)$, of which the inputs are $\textbf{I}^t$, $\textbf{a}^t$ and $\textbf{o}^t$, and the output is $\textbf{I}^{t+1}$.

    \item  \underline{\emph{Reward:}} The expected immediate reward at belief state $\textbf{I}^t$ is defined as the weighted sum of the instantaneous Aol of all end nodes, i.e., $R(\textbf{I}^t) \triangleq \sum_{i=1}^K \omega_i h_i^t$.
    Then, the EWSAoI in \eqref{EWSAoI} can be evaluated by
    \begin{equation} \label{eq:EWSAoI-po}
    \frac{1}{TK} \mathbb{E}\left[\left.\sum_{t=1}^{T} R(\textbf{I}^t) \right| \textbf{I}^1, \bm{\pi} \right],
    \end{equation}
    where $\textbf{I}^1$ is a given initial belief state.

    \item \underline{\emph{Policy:}} In the above equation, $\bm{\pi}$ is a given policy defined as $\bm{\pi} = [ \pi^1,\pi^2,\ldots,\pi^T ]$, where $\pi^t$ is a mapping from the belief space $\bm{\mathcal{I}}$ to the action space $\bm{\mathcal{A}}$, i.e., decides which action $\textbf{a}^t$ should be taken when the POMDP is in belief state $\textbf{I}^t$.
    Our aim is to find the optimal policy that minimizes the average reward in \eqref{eq:EWSAoI-po}, i.e.,
    \begin{equation}
    \bm{\pi}^\ast = \mathop{\arg\min}_{\bm{\pi}} \frac{1}{TK} \mathbb{E}\left[\left.\sum_{t=1}^{T} R(\textbf{I}^t) \right| \textbf{I}^1, \bm{\pi} \right].
    \end{equation}
\end{enumerate}

\begin{figure}
	\centering
	\includegraphics[width=3.5in]{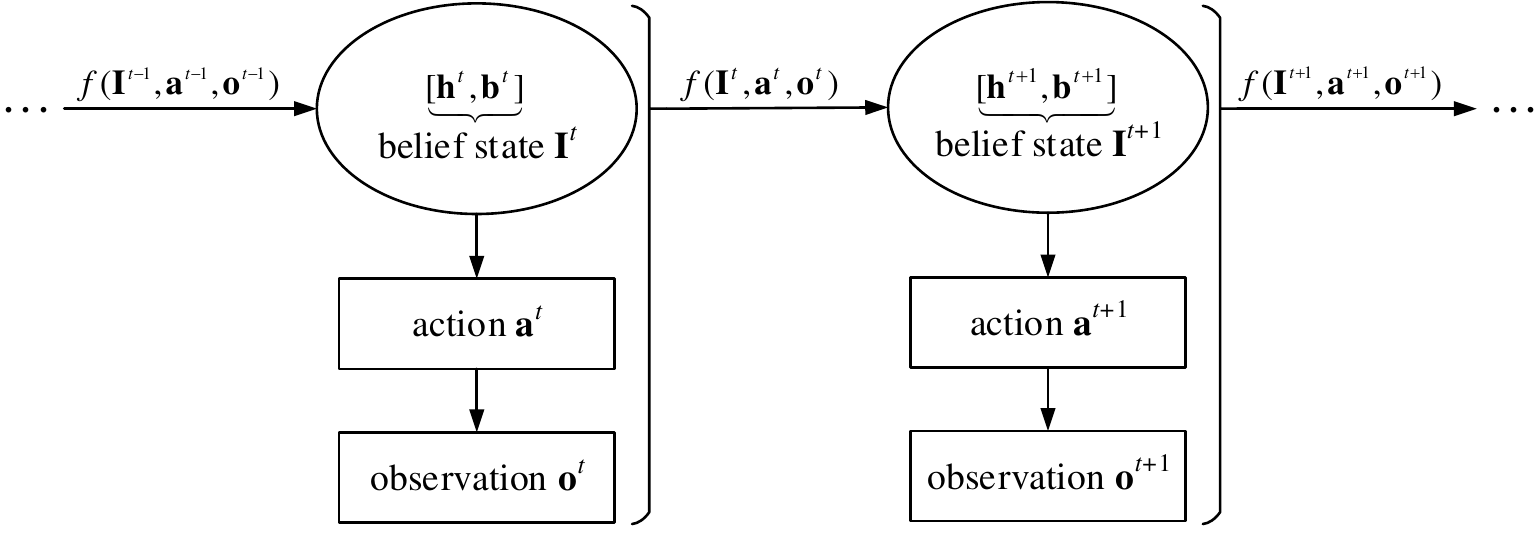}
	\caption{An illustration of belief states, actions, observations and the update of belief states.}
\end{figure}

To illustrate the POMDP formulation, we depict its belief states, actions, observations and update of belief states in Fig. 2. 

\section{Policy Design for the Formulated POMDP}

In this section, we first propose a DP algorithm to find the optimal policy of the formulated POMDP and then devise a myopic policy with low-complexity and near-optimal performance.

\subsection{Dynamic Programming for the Optimal Policy}
We follow \cite{Ahmad2009Optimality} and resort to the DP framework for finding the optimal policy of the POMDP formulated in Section III.
The DP method solves complex problems by breaking them down into a sequence of simpler sub-problems and then recursively combining solutions of sub-problems. It is worth mentioning that the space of belief states of the POMDP is countable for any given initial belief state $\textbf{I}^1$. We denote the finite set of belief states in time slot $t$ by $\bm{\mathcal{I}}^t$. The expected total reward of $\forall \textbf{I}^t \in \bm{\mathcal{I}}^t$ can be denoted by the inner product $\textbf{b}^t\cdot\textbf{V}(\textbf{I}^t)$, where $\textbf{V}(\textbf{I}^t)\triangleq [v_{\textbf{I}^t}(\textbf{z}_1^t),v_{\textbf{I}^t}(\textbf{z}_2^t),\ldots,v_{\textbf{I}^t}(\textbf{z}_{|\bm{\mathcal{Z}}|}^t)]$ is a $|\bm{\mathcal{Z}}|$ dimensional vector by recalling that the update of $\textbf{h}^t$ is deterministic.
Note that $\textbf{b}^t\cdot\textbf{V}(\textbf{I}^t)$ incorporates rewards from time slot $t$ onward.
The DP algorithm is formally described as follows.

\begin{algorithm}
    \caption{The DP Algorithm}
  \begin{algorithmic}[1]
        \STATE \textbf{Initialization:}\\
        Set $\textbf{V}^\ast (\textbf{I}^T)=R(\textbf{I}^T)\textbf{1}_{|\bm{\mathcal{Z}}|}$ for $\forall \textbf{I}^T \in \bm{\mathcal{I}}^T$, where $\textbf{1}_{|\bm{\mathcal{Z}}|}$ is a $|\bm{\mathcal{Z}}|$ dimensional vector with all entries equal to $1$, and $t=T-1$.
        \STATE \textbf{Backward Induction:}\\
        1) For $\forall \textbf{I}^t \in \bm{\mathcal{I}}^t$, compute $\pi^\ast (\textbf{I}^t)$.
        \begin{equation} \label{eq:dp-policy}
        \begin{aligned}
        &\pi^\ast(\textbf{I}^t)=\mathop{\arg\min}_{\textbf{a}^t \in \bm{\mathcal{A}}}
        \sum_{\textbf{z}^t \in \bm{\mathcal{Z}}} b^t(\textbf{z}^t) \bigg[ R(\textbf{I}^t) +
        \sum_{\textbf{o}^t \in \bm{\mathcal{O}}} \vartheta\times \\ &\text{Pr}(\textbf{o}^t|\textbf{z}^t,\textbf{a}^t)
        \sum_{\textbf{z}^{t+1} \in \bm{\mathcal{Z}}} \text{Pr}(\textbf{z}^{t+1}|\textbf{z}^t) v^\ast_{\textbf{I}^{t+1}} (\textbf{z}^{t+1}) \bigg],
        \end{aligned}
        \end{equation}
        $\forall \textbf{I}^{t+1} \in \bm{\mathcal{I}}^{t+1}$, where $\vartheta=1$ if $f(\textbf{I}^t,\textbf{a}^t,\textbf{o}^t)=\textbf{I}^{t+1}$, and $\vartheta=0$ otherwise.\\
        2) For $\forall \textbf{I}^t \in \bm{\mathcal{I}}^t$, compute $v^\ast_{\textbf{I}^t} (\textbf{z}^{t})$ for $\forall \textbf{z}^{t} \in \bm{\mathcal{Z}}$.
        \begin{equation} \label{eq:dp-evaluation}
        \begin{aligned}
        &v^\ast_{\textbf{I}^t} (\textbf{z}^{t}) = R(\textbf{I}^t) +
        \sum_{\textbf{o}^t \in \bm{\mathcal{O}}} \vartheta\,\text{Pr}(\textbf{o}^t|\textbf{z}^t,\pi^\ast (\textbf{I}^t))\times\\
        &\sum_{\textbf{z}^{t+1} \in \bm{\mathcal{Z}}} \text{Pr}(\textbf{z}^{t+1}|\textbf{z}^t) v^\ast_{\textbf{I}^{t+1}} (\textbf{z}^{t+1}),\,\,\forall \textbf{I}^{t+1} \in \bm{\mathcal{I}}^{t+1}.
        \end{aligned}
        \end{equation}
        \STATE \textbf{Stopping Rule:}\\
        If $t=1$, stop. Otherwise, set $t=t-1$ and go to step 2.
    \end{algorithmic}
\end{algorithm}

The recursion simplifies the evaluation and optimization of { $\textbf{V}^\ast (\textbf{I}^1)$} over $T$ time slots into a sequence of $T-1$ one-step computations.
As shown in \eqref{eq:dp-evaluation}, in each step, the value of $v^\ast_{\textbf{I}^t} (\textbf{z}^{t})$ equals the immediate reward plus the expected total reward over the remaining time slots.
The optimal policy $\bm{\pi}^\ast$ is defined as { $\pi^{\ast,t}: \textbf{I}^t \to \pi^\ast (\textbf{I}^t)$} for { $\forall \textbf{I}^t \in \bm{\mathcal{I}}^t$}.
In particular, { $(\textbf{b}^1\cdot\textbf{V}^\ast (\textbf{I}^1))/(TK)$} is the minimal EWSAoI given the initial belief state $\textbf{I}^1$.

The DP algorithm represents an effective solution to find the optimal policy.
However, the recursion is computationally prohibitive due to the following reasons.
First, the AoI and local age tends to be large in real systems.
Second, the dimension of the probability distribution $\textbf{b}^t$ grows exponentially with the number of end nodes.
Thus, it is crucial to find a low-complexity and near-optimal policy.

\subsection{A Myopic Policy}
In our problem, the local age of different end nodes evolves independently as described in Section II.
As such, the monitor can only maintain probability distributions of the local age of each end node, which are sufficient statistics for the POMDP.
We let $\mathbb{B}^t=[\textbf{b}_1^t,\textbf{b}_2^t\ldots,\textbf{b}_K^t ]$ denote these distributions, where $\textbf{b}_i^t$ is the probability distribution of the local age of end node $i$.
Let $b_i^t(z_i^t)$ denote the probability assigned to local age $z_i^t$ by distribution $\textbf{b}_i^t$, satisfying $b_i^t(z_i^t) \in [0,1]$ for all $z_i^t \in \bm{{\cal{T}}}$, and $\sum_{z_i^t \in \bm{{\cal{T}}}} b_i^t(z_i^t)=1$.
Then, the belief state of the POMDP can be expressed as $\mathbb{I}^t \triangleq [ \textbf{h}^t, \mathbb{B}^t ]$.

We then propose a myopic policy that minimizes the expected reward of the next time slot, also known as a one-step expected reward.
Given $\mathbb{I}^t$ for the POMDP, if action $\textbf{a}^t$ is chosen in time slot $t$, the one-step expected reward of the system is given by
\begin{equation} \label{eq:one-step}
\begin{aligned}
&\hat{R}(\mathbb{I}^t,\textbf{a}^t) = \sum_{i=1}^K \omega_i \bigg[ (1-a_i^t)(h_i^t+1)+ \\
&a_i^t \bigg(p_i \sum_{z_i^t \in \bm{{\cal{T}}}} b_i^t(z_i^t)(z_i^t+1) + (1 - p_i) (h_i^t+1)\bigg) \bigg].
\end{aligned}
\end{equation}
For a scheduled end node, its expected AoI in next time slot is $\sum_{z_i^t \in \bm{{\cal{T}}}} b_i^t(z_i^t)(z_i^t+1)$ with probability $p_i$ or $(h_i^t+1)$ with probability $(1 - p_i)$.
For an arbitrary unscheduled end node, its expected AoI in the next time slot is $(h_i^t+1)$.
The myopic policy for belief state $\mathbb{I}^t$ can be obtained by
\begin{equation} \label{eq:myopic}
\begin{aligned}
\tilde{\pi}^\ast & (\mathbb{I}^t) = \mathop{\arg\min}_{\textbf{a}^t \in \bm{\mathcal{A}}} \hat{R}(\mathbb{I}^t,\textbf{a}^t).
\end{aligned}
\end{equation}
Then, the myopic policy $\tilde{\bm\pi}^\ast$ is defined as $\tilde{\pi}^{\ast,t}: \mathbb{I}^t \to \tilde{\pi}^\ast (\mathbb{I}^t)$.
Compared with the optimal policy, the myopic policy is easier to implement. Not only $\mathbb{B}^t$ reduces the dimension of $\textbf{b}^t$ from $|\bm{\mathcal{T}}|^K$ to $K|\bm{\mathcal{T}}|$, growing linearly with the number of end nodes, but also the myopic policy only relies on the one-step expected reward instead of the expected total reward. The proposed myopic policy is formally described in Algorithm 2.

\begin{algorithm}
	\caption{The Myopic Policy}
	\begin{algorithmic}[1]
		\STATE \textbf{Initialization:}\\
		Set $t=1$ and give initial belief state $\mathbb{I}^1$. For each end node, the monitor maintains its AoI $h_i^t$ and a probability distribution $\textbf{b}_i^t$ of its local age.
		\STATE \textbf{Obtain the Myopic Policy:}\\
		In time slot $t$, the monitor chooses action $\tilde{\textbf{a}}^{t}$ by 
		\[
		\begin{aligned}
		\tilde{\textbf{a}}^{t} = \mathop{\arg\min}_{\textbf{a}^t \in \bm{\mathcal{A}}} \hat{R}(\mathbb{I}^t,\textbf{a}^t).
		\end{aligned}
		\]
		\STATE \textbf{Update the Belief State:}\\
		After taking action $\tilde{\textbf{a}}^{t}$, the monitor makes observation $\textbf{o}^t$.
		For each end node, the monitor updates its $h_i^{t+1}$ by \eqref{eq:h-update}.
        If $\hat{z}_i^{t} \neq X$, the monitor updates its $\textbf{b}_i^{t+1}$ by 
        \[
        b_i^{t+1}(z_i^{t+1})=
        \begin{cases}
        \lambda_i, & \text{if}\,\,z_i^{t+1} = 1, \\
        1-\lambda_i, & \text{if}\,\,z_i^{t+1} = \hat{z}_i^{t}+1, \\
        0, & \text{otherwise}.
        \end{cases}
        \]
        Otherwise, the monitor updates its $\textbf{b}_i^{t+1}$ by
        \[
        b_i^{t+1}(z_i^{t+1})=\sum_{z_i^t \in \bm{\mathcal{T}}} \text{Pr} (z_i^{t+1}|z_i^t) b_i^{t}(z_i^t).
        \]
		\STATE \textbf{Stopping Rule:}\\
		If $t=T$, stop. Otherwise, set $t=t+1$ and go to step 2.
	\end{algorithmic}
\end{algorithm}

\section{Performance Evaluation}

In this section, after introducing a physical-layer model to obtain successful transmission probability $p_i$, we evaluate the DP algorithm and myopic policy via simulations.

\subsection{Successful Transmission Probability}
In time slot $t$, the small-scale fading from end node $i$ to the monitor is denoted by $g_i^t$, which is assumed to follow an exponential random variable with a unit mean. The large-scale fading is denoted by $d_i^{-\tau}$, where $d_i$ is the distance from end node $i$ to the monitor and $\tau$ is the path-loss factor. The additive white Gaussian noise follows a complex Gaussian distribution ${\cal{CN}}(0,\sigma^2)$.
The achievable rate is computed by $r_i = \log_2\left({d_i^{-\tau}g_i^t P}/{\sigma^2} + 1\right)$, where $P$ is the transmit power, and the signal-to-noise ratio (SNR) is calculated by $\text{SNR} = d_i^{-\tau}P/\sigma^2$.
If $r_i$ is below the threshold $r_{th}$, the transmission of end node $i$ is deemed to be unsuccessful. Consequently, the successful transmission probability of end node $i$ can be obtained by
\begin{equation}
\begin{aligned}
p_i &= 1 - \text{Pr}\left(r_i < r_{th} \right) = 1 - \text{Pr}\left( g_i^t < \sigma^2d_i^\tau\frac{2^{r_{th}} - 1}{P} \right) \\
&\overset{(a)}{=} \exp \left(-  \frac{\sigma^2d_i^\tau(2^{r_{th}} - 1)}{P} \right)
\end{aligned}
,\end{equation}
where step $(a)$ follows by the cumulative density function of the exponential random variable with a unit mean.
We set the distance $d_i=5\text{m}$, the path-loss factor $\tau = 2$, the threshold $r_{th}= 1\,\text{bps}/\text{Hz}$ for all simulation runs in this section.

\subsection{Comparisons with Simulation Results}

In solving the POMDP-based policies, a state truncation $D$ is applied to approximate the countable state space, i.e., the AoI and local age are both upper bounded by $D$ for $\forall i,t$.

Fig. 3 shows the analytical and simulation results of the optimal and myopic policies as a function of the SNR, in which we set $K=2$, $T=25$, $D=8$, $\lambda_i=0.4$ and $\omega_i= 1$ for $\forall i \in \bm{{\cal{K}}}$.
Each simulation result is obtained from $10^6$ independent simulation runs.
For both policies, it is shown that the analytical and simulation results are well matched, which verifies the accuracy of the POMDP formulation.

Fig. 4 compares the analytical results of the myopic policy with that of the optimal policy, in which the parameter setting is the same as that in Fig. 3 except $D=4,6,8,10$. It is shown that the performance of the myopic policy can approach that of the optimal policy, i.e., the myopic policy can achieve near-optimal performance.
Meanwhile, the curves indicate that the increase of the EWSAoI becomes slower with the increase of the state truncation $D$.
This is because the AoI and local age of end nodes tend to be limited when the monitor takes the optimal or myopic policy aiming to minimize the EWSAoI.

\begin{figure}
	\centering
	\includegraphics[width=2.8in]{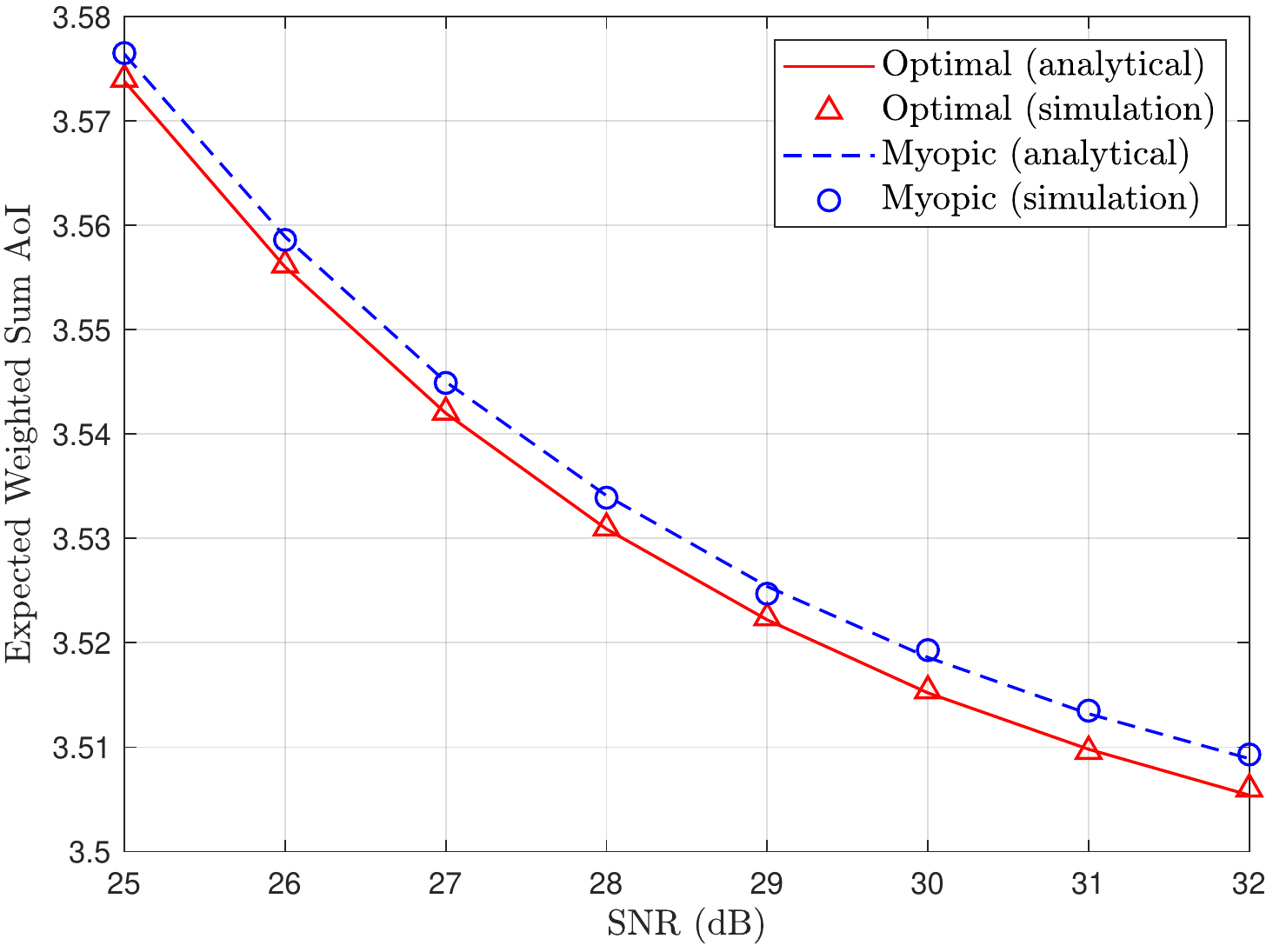}
	\caption{Analytical results v.s. simulation results, $K=2,D=8$.}
\end{figure}

\begin{figure}
	\centering
	\includegraphics[width=2.8in]{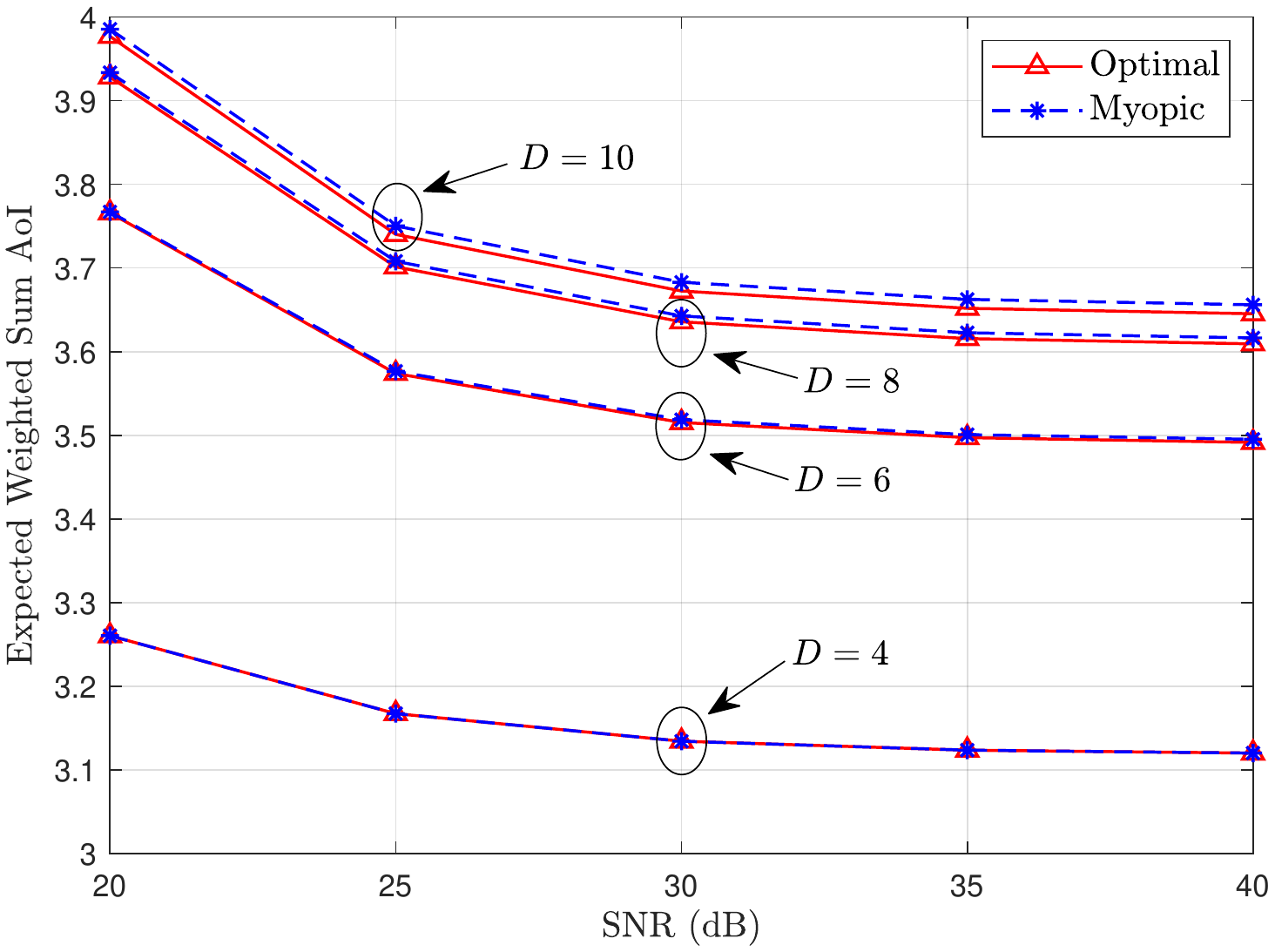}
	\caption{Proposed myopic policy v.s. optimal policy, $K=2$.}
\end{figure}

\subsection{Comparison with Baseline Policies}

We compare the proposed myopic policy with two baseline policies, described as follows:

\begin{enumerate}
    \item \emph{MDP Policy:} We introduce a myopic policy proposed in \cite{13}, which assumes a complete knowledge of status update arrivals. Denote by ``MDP" this myopic policy.
    \item \emph{MaxAoI Policy:} We propose a myopic policy, which assumes no knowledge of status update arrivals.
    Specifically, the monitor always schedules the end node with maximum AoI to transmit.
    The policy only relies on the fully observable AoI available at the monitor.
    Denote by ``MaxAoI" this myopic policy.
\end{enumerate}

Fig. 5 shows the simulation results of the myopic, MaxAoI and MDP policies as a function of the SNR, in which we set $K=2$, $T=10^6$, $D=30$, $\lambda_i=0.4$ and $\omega_i= 1$ for $\forall i \in \bm{{\cal{K}}}$.
Each simulation result is obtained from $10$ independent simulation runs.
Since the MDP policy has a complete knowledge of status update arrivals, there is a gap between the myopic and MDP policies.
The MaxAoI policy utilizes no knowledge of status update arrivals, thus it has the worst performance among three policies.
It is further shown that the gap between the myopic and MDP policies becomes larger as the number of end nodes increases.
This is because the gain resulting from fully observable status update arrivals is augmented as the number of end nodes increases.

\begin{figure}
    \centering
    \includegraphics[width=2.8in]{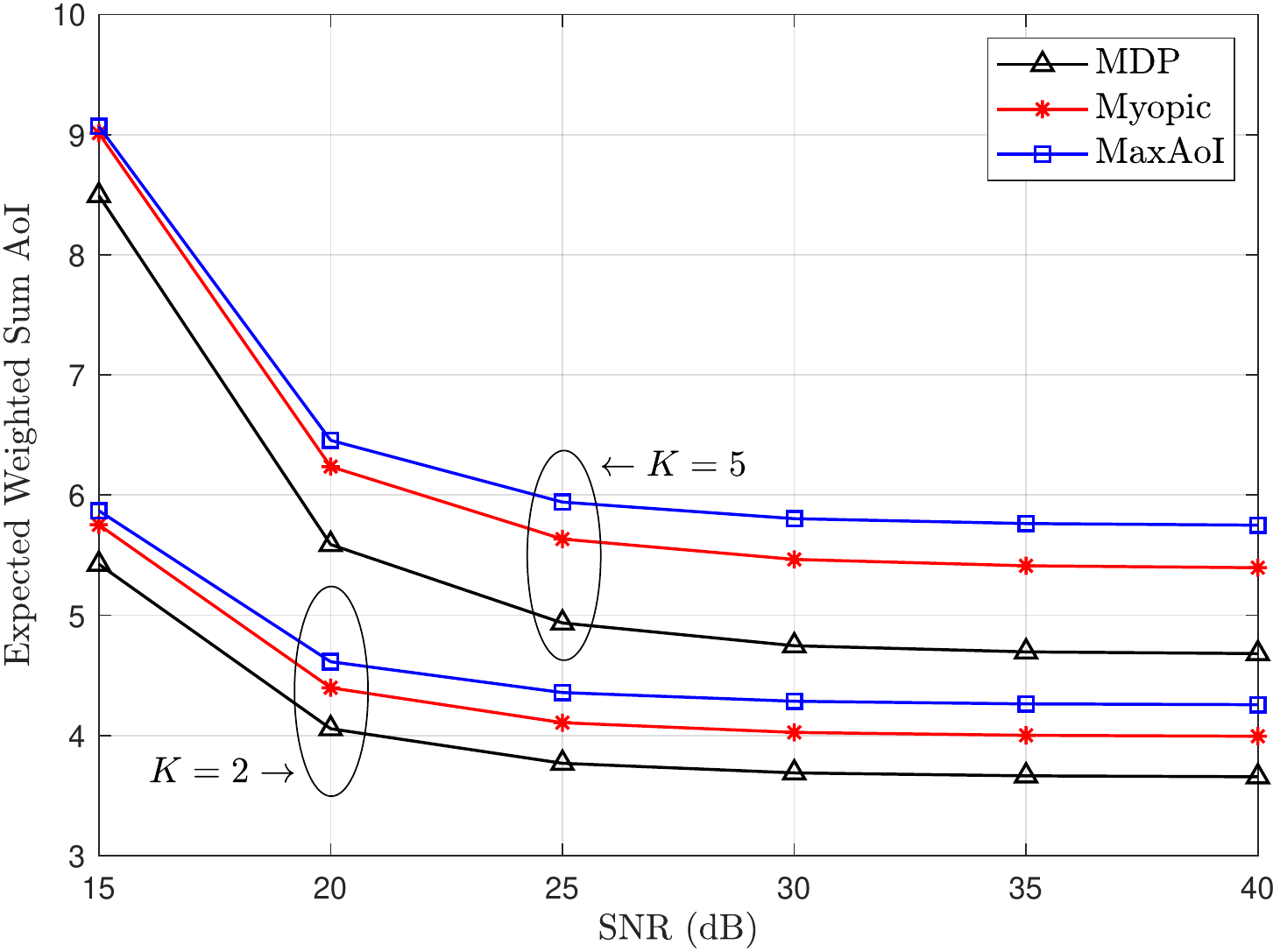}
    \caption{Proposed myopic policy v.s. baseline policies, $D=30$.}
\end{figure}

Fig. 6 shows the simulation results of the myopic, MaxAoI and MDP policies as a function of the packet arrival rate, in which the parameter setting is the same as that in Fig. 5 except $\text{SNR}=30\text{dB}$.
It is shown that as the status update arrival rate increases, the performance of these three polices turns to converge. This is because, for large status update arrival rates, the importance of status update arrival knowledge is marginal when minimizing the EWSAoI. For the extreme case with $\lambda_i = 1$ for $\forall i \in \bm{{\cal{K}}}$, the considered system will be equivalent to the ``generate-at-will" model, where there is no uncertainty on the status update arrival knowledge.

\begin{figure}
	\centering
	\includegraphics[width=2.8in]{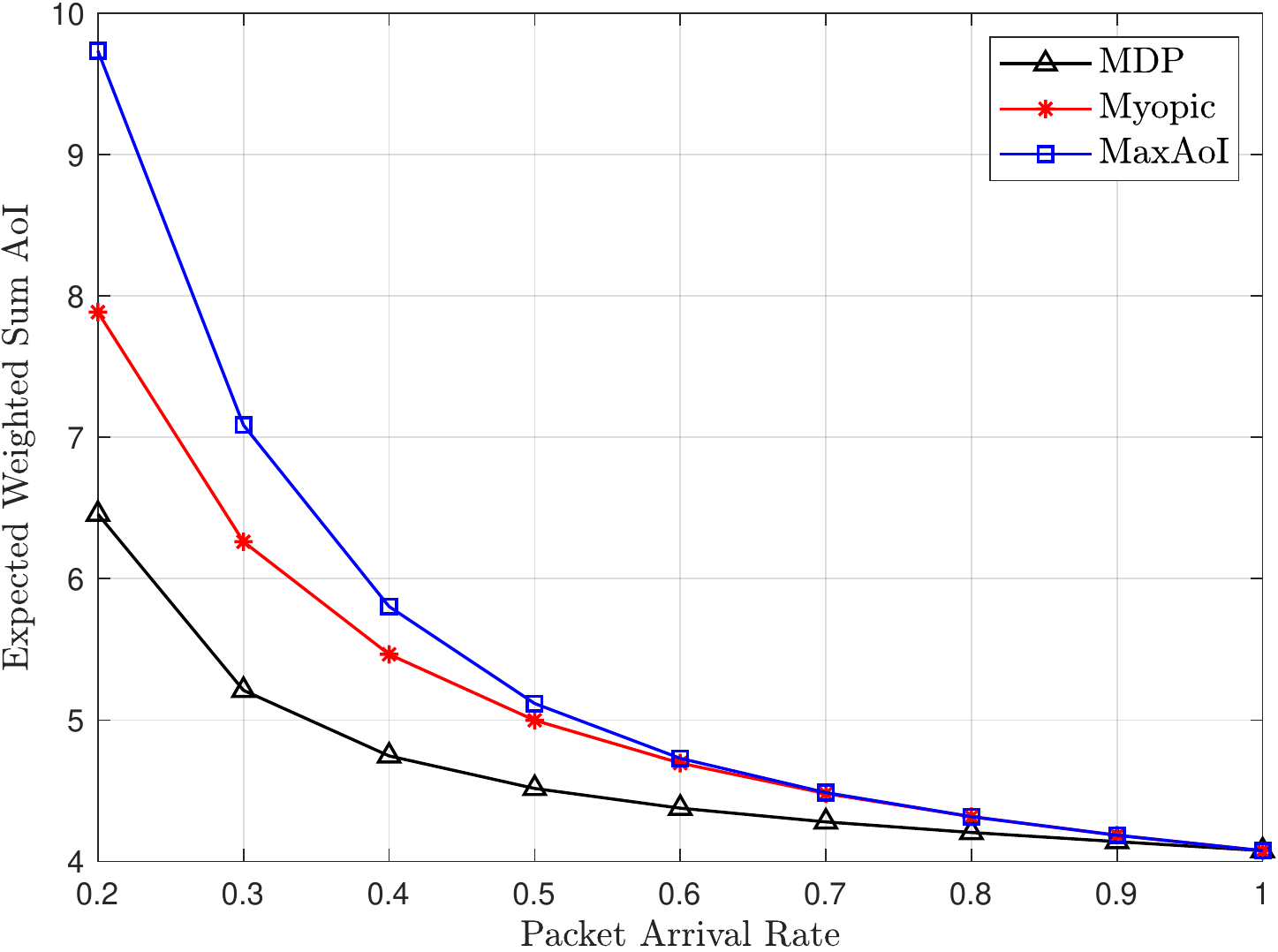}
	\caption{Proposed myopic policy v.s. baseline policies, $D=30,K=5$.}
\end{figure}

\section{Conclusions}

In this paper, we have investigated the information-freshness-oriented scheduling problem in the multiuser uplink system, where the monitor has a partial knowledge of status update arrivals at the end node side.
To tackle this problem, a POMDP has been formulated to characterize the dynamic behavior of such system.
A DP algorithm has been developed to achieve the optimal policy, and a myopic policy with low-complexity and near-optimal performance has been further proposed.
Simulation results have shown that the performance of the myopic policy approaches that of the optimal policy, and is better than that of the baseline policy which utilizes no knowledge of status update arrivals. Moreover, simulation results have indicated that the role of status update arrival knowledge in minimizing the network-wide AoI becomes insignificant when the status update arrival rate goes large.

\bibliographystyle{IEEEtran}
\bibliography{POMDP_GC}

\begin{thebibliography}{10}
\providecommand{\url}[1]{#1}
\csname url@samestyle\endcsname
\providecommand{\newblock}{\relax}
\providecommand{\bibinfo}[2]{#2}
\providecommand{\BIBentrySTDinterwordspacing}{\spaceskip=0pt\relax}
\providecommand{\BIBentryALTinterwordstretchfactor}{4}
\providecommand{\BIBentryALTinterwordspacing}{\spaceskip=\fontdimen2\font plus
\BIBentryALTinterwordstretchfactor\fontdimen3\font minus
  \fontdimen4\font\relax}
\providecommand{\BIBforeignlanguage}[2]{{%
\expandafter\ifx\csname l@#1\endcsname\relax
\typeout{** WARNING: IEEEtran.bst: No hyphenation pattern has been}%
\typeout{** loaded for the language `#1'. Using the pattern for}%
\typeout{** the default language instead.}%
\else
\language=\csname l@#1\endcsname
\fi
#2}}
\providecommand{\BIBdecl}{\relax}
\BIBdecl

\bibitem{000}
A.~Kosta, N.~Pappas, and V.~Angelakis, ``Age of information: A new concept,
  metric, and tool,'' \emph{Foundations and Trends in Netw.}, vol.~12, no.~3,
  pp. 162--259, 2017.

\bibitem{001}
Y.~Sun, I.~Kadota, R.~Talak, and E.~Modiano, ``Age of information: {A} new
  metric for information freshness,'' \emph{Synthesis Lectures on Commun.
  Networks}, vol.~12, no.~2, pp. 1--224, 2019.

\bibitem{-1}
S.~{Kaul}, M.~{Gruteser}, V.~{Rai}, and J.~{Kenney}, ``Minimizing age of
  information in vehicular networks,'' in \emph{Proc. Annu. IEEE Commun. Soc.
  Conf. Sensor, Mesh Ad-Hoc Commun. Netw. (SECOM)}, 2011, pp. 350--358.

\bibitem{1}
S.~{Kaul}, R.~{Yates}, and M.~{Gruteser}, ``Real-time status: {How} often
  should one update?'' in \emph{Proc. IEEE Conf. Comput. Commun. (INFOCOM)},
  2012, pp. 2731--2735.

\bibitem{2}
M.~{Costa}, M.~{Codreanu}, and A.~{Ephremides}, ``On the age of information in
  status update systems with packet management,'' \emph{IEEE Trans. Infor.
  Theory}, vol.~62, no.~4, pp. 1897--1910, 2016.

\bibitem{3}
C.~{Kam}, S.~{Kompella}, and A.~{Ephremides}, ``Age of information under random
  updates,'' in \emph{Proc. IEEE Int. Symp. Inf. Theory (ISIT)}, 2013, pp.
  66--70.

\bibitem{4}
Y.~{Sun}, E.~{Uysal-Biyikoglu}, R.~D. {Yates}, C.~E. {Koksal}, and N.~B.
  {Shroff}, ``Update or wait: How to keep your data fresh,'' \emph{IEEE Trans.
  Infor. Theory}, vol.~63, no.~11, pp. 7492--7508, 2017.

\bibitem{gu2019timely}
Y.~Gu, H.~Chen, Y.~Zhou, Y.~Li, and B.~Vucetic, ``Timely status update in
  internet of things monitoring systems: An age-energy tradeoff,'' \emph{IEEE
  Internet of Things Journal}, 2019.

\bibitem{gu2019Minimizing}
Y.~{Gu}, H.~{Chen}, C.~{Zhai}, Y.~{Li}, and B.~{Vucetic}, ``Minimizing age of
  information in cognitive radio-based iot systems: Underlay or overlay?''
  \emph{IEEE Internet of Things Journal}, vol.~6, no.~6, pp. 10\,273--10\,288,
  2019.

\bibitem{wang2019minimizing2}
Q.~Wang, H.~Chen, Y.~Gu, Y.~Li, and B.~Vucetic, ``Minimizing the age of
  information of cognitive radio-based iot systems under a collision
  constraint,'' \emph{arXiv preprint arXiv:2001.02482}, 2020.

\bibitem{5}
I.~{Kadota}, A.~{Sinha}, E.~{Uysal-Biyikoglu}, R.~{Singh}, and E.~{Modiano},
  ``Scheduling policies for minimizing age of information in broadcast wireless
  networks,'' \emph{IEEE/ACM Trans. Netw.}, vol.~26, no.~6, pp. 2637--2650,
  2018.

\bibitem{wang2020minimizing}
Q.~Wang, H.~Chen, Y.~Li, and B.~Vucetic, ``Minimizing age of information via
  hybrid {NOMA/OMA},'' \emph{arXiv preprint arXiv:2001.04042}, 2020.

\bibitem{8}
I.~{Kadota} and E.~{Modiano}, ``Minimizing the age of information in wireless
  networks with stochastic arrivals,'' \emph{IEEE Trans. Mobile Comput.}, pp.
  1--1, 2019.

\bibitem{9}
Y.~{Hsu}, E.~{Modiano}, and L.~{Duan}, ``Scheduling algorithms for minimizing
  age of information in wireless broadcast networks with random arrivals,''
  \emph{IEEE Trans. Mobile Comput.}, pp. 1--1, 2019.

\bibitem{11}
Z.~{Jiang}, B.~{Krishnamachari}, X.~{Zheng}, S.~{Zhou}, and Z.~{Niu}, ``Timely
  status update in wireless uplinks: Analytical solutions with asymptotic
  optimality,'' \emph{IEEE Internet Things J.}, vol.~6, no.~2, pp. 3885--3898,
  2019.

\bibitem{12}
J.~{Sun}, Z.~{Jiang}, B.~{Krishnamachari}, S.~{Zhou}, and Z.~{Niu},
  ``Closed-form whittle’s index-enabled random access for timely status
  update,'' \emph{IEEE Trans. Commun.}, vol.~68, no.~3, pp. 1538--1551, 2020.

\bibitem{13}
H.~Chen, Q.~Wang, Z.~Dong, and N.~Zhang, ``Multiuser scheduling for minimizing
  age of information in uplink {MIMO} systems,'' \emph{arXiv preprint
  arXiv:2002.00403}, 2020.

\bibitem{Ahmad2009Optimality}
S.~H.~A. {Ahmad}, M.~{Liu}, T.~{Javidi}, Q.~{Zhao}, and B.~{Krishnamachari},
  ``Optimality of myopic sensing in multichannel opportunistic access,''
  \emph{IEEE Trans. Inf. Theory}, vol.~55, no.~9, pp. 4040--4050, 2009.

\end{thebibliography}

\end{document}